\documentclass[%
 reprint,
superscriptaddress,
 amsmath,amssymb,
longbibliography,
]{revtex4-1}

\usepackage{import} 
\usepackage{array}
\usepackage{calc}
\usepackage[normalem]{ulem}

\usepackage{etoolbox,siunitx}
  \robustify\bfseries

  \usepackage{grffile}
  \usepackage{graphicx}
  \usepackage{color}
  \usepackage{dcolumn}% Align table columns on decimal point
  \usepackage{bm}% bold math
\usepackage{multirow}
\usepackage{pbox}
\usepackage{tabularx,booktabs} 
\usepackage{hyperref} % to include url easier in tex 

\definecolor{pink}{rgb}{0.858, 0.188, 0.478}
\definecolor{green}{rgb}{0, 0.4, 0}

\makeatletter
\def\NAT@def@citea{\def\@citea{\NAT@separator}}
\makeatother

\begin{document}
  %\preprint{APS/123-QED}


\author{Kuang-Chung Wang}
\affiliation{School of Electrical and Computer Engineering, Purdue University, West Lafayette, IN 47906, USA}

\author{Roberto Grassi}
\affiliation{Silvaco Co., Santa Clara, CA , USA}

\author{Yuanchen Chu}
\affiliation{School of Electrical and Computer Engineering, Purdue University, West Lafayette, IN 47906, USA}
\author{Shree Hari Sureshbabu }%
\affiliation{School of Electrical and Computer Engineering, Purdue University, West Lafayette, IN 47906, USA}

\author{Junzhe  Geng}%
\affiliation{School of Electrical and Computer Engineering, Purdue University, West Lafayette, IN 47906, USA}

\author{Prasad Sarangapani}
\affiliation{School of Electrical and Computer Engineering, Purdue University, West Lafayette, IN 47906, USA}

\author{Xinchen Guo}%
\affiliation{School of Electrical and Computer Engineering, Purdue University, West Lafayette, IN 47906, USA}

\author{Mark Townsend}
\affiliation{Silvaco Co., Santa Clara, CA , USA}

\author{Tillmann Kubis}
\affiliation{School of Electrical and Computer Engineering, Purdue University, West Lafayette, IN 47906, USA}
\affiliation{Network for Computational Nanotechnology, Purdue University, West Lafayette, IN 47906, USA}
\affiliation{Purdue Center for Predictive Materials and Devices, Purdue University,  West Lafayette, IN 47906, USA}
\affiliation{Purdue Institute of Inflammation, Immunology and Infectious Disease,  West Lafayette, IN 47906, USA}


\date{\today}% It is always \today, today,

\begin{abstract}
State-of-the-art industrial semiconductor device modeling is based on highly efficient Drift-Diffusion (DD) models that include some quantum corrections for  nanodevices.
In contrast, latest academic quantum transport models are based on the non-equilibrium Green's function (NEGF) method that cover all coherent and incoherent quantum effects consistently.
Carrier recombination and generation in optoelectronic nanodevices represent an immense numerical challenge when solved within NEGF. 
In this work, the numerically efficient B\"uttiker-probe model is expanded to include electron-hole recombination and generation in the NEGF framework. 
Benchmarks of the new multiple-particle B\"{u}ttiker probe method against state-of-the-art quantum-corrected DD models show quantitative agreements except in cases of pronounced tunneling and interference effects.
\end{abstract}

\title{Introduction of Multi-particle B\"{u}ttiker Probes - Bridging the Gap between Drift Diffusion and Quantum Transport}
\maketitle

\section{Introduction}

State-of-the-art semiconductor device fabrication techniques allow for device design at the atomistic length scale~\cite{Pang2019}.
The performance of nanodevices is equally influenced by coherent quantum mechanical phenomena (such as confinement, tunneling and interference)~\cite{Krogstrup2013,Ilatikhameneh2018} and incoherent scattering of electrons on device imperfections and lattice vibrations~\cite{Luisier2012,Charles2016}. 
The performance of solar cells~\cite{Burgelman2000,Wang2011}, lasers~\cite{Jirauschek2014} and light emitting diodes~\cite{Geng2018,Pimputkar2009} critically depend on the incoherent interaction of electrons with phonons and photons and the interplay between radiative, Auger and Shockley-Read-Hall (SRH) recombination.
Carrier generation and recombination affect the off-current and switching characteristics of tunneling field effect transistors~\cite{Ghedini2012,Smets2017}.
The non-equilibrium Green's function method (NEGF) is among the most general methods to describe coherent and incoherent transport physics~\cite{Charles2016}.
Due to the large numerical load when incoherent scattering is included in the self-consistent Born approximation~\cite{Charles2016}, NEGF is typically applied in the coherent transport limit~\cite{Svizhenko2002}.
This is particularly problematic in nanodevices with pronounced incoherent effects~\cite{Klimeck1995}.
There are various algorithms to include incoherent scattering in NEGF. 
The self-consistent Born approximation can rigorously treat incoherent scattering~\cite{Charles2016}, carrier generation~\cite{Aeberhard2018} and recombination~\cite{Aeberhard2019}. 
However, the self-consistent Born approximation involves several nonlinear and highly dimensional integro-differential equations, which yield high computational load. 
The multi-scale and multi-physics NEGF implementation of Ref.~\onlinecite{Geng2018} had been designed for modeling light emitting diodes with low numerical load. 
It requires full charge carrier thermalization in each quantum well.
Electron-hole recombination process is limited to the fully thermalized quantum wells as well. 
The B\"{u}ttiker-probe model~\cite{Venugopal2003b,Greck2015,Witzigmann} represents a good compromise between the accuracy of NEGF and the numerical efficiency of heuristic scattering models for devices with incomplete carrier thermalization. 
In this work, the B\"{u}ttiker-probes are extended to cover electron-hole recombination and generation in addition to its traditional application space of mobility limiting intraband scattering. 
Current conservation for intraband and interband scattering is ensured.
NEGF predictions with the augmented B\"{u}ttiker-probes are benchmarked against the drift-diffusion(DD) method of Atlas~\cite{atlasweb}. 
DD is at the core of industrial technology computer aided design (TCAD) tools for micro-scale devices~\cite{Kim2007,Wu2012}.
DD is known for its computational efficiency, but it requires to additional correction terms for mimicking coherent quantum effects~\cite{Bandyopadhyay1987,Bufler2004,Procel2019}. 
The NEGF with B\"{u}ttiker-probe transport predictions of pn-junctions agree quantitatively with DD results of the ON-current density and with results of the density in thermalized device regions.
pn junctions that include quantum wells can serve as solar cells~\cite{Lang2012b,Wu2003a} and photo-detectors~\cite{Jain2018}.
Therefore, the new method is also benchmarked against a quantum corrected DD model for carrier recombination and light absorption.
Deviations are found in cases with pronounced tunneling and interference effects.
It is worth to mention the method is compatible with arbitrary basis representations, ranging from effective mass\cite{Wang2004} and k.p~\cite{Huang2015c} to  atomistic approaches~\cite{Wang2017,Valencia2018a,Zahid2012,Chen2018}.

\section{Method}
The new B\"{u}ttiker probes are benchmark against state-of-the-art TCAD methods on two devices - a GaN pn diode and a GaN pn diode including an intrinsic InGaN quantum well centered at the p-n interface.
%The pn diode is composed of 10~nm p type and 10~nm n type doped GaN with the doping concentration of $10^{20}/cm^3$ in each region.
%Both regions are periodic in the transverse directions. 
%The pn diode with the quantum well differs from the pn diode by a 3.0~nm thick intrinsic In$_{0.13}$Ga$_{0.87}$N quantum well layer. 
GaN electrons and holes are modeled each in effective mass assuming the isotropic masses $m_e=0.2m^*$ and $m_h=1.25m^*$\cite{Pugh1999}. 
The position dependent electron and hole recombination and generation~\cite{Piprek2010} rate ($R_{R/G}$) depends on the contributions of SRH, radiative recombination, Auger effect and light absorption
\begin{equation}\label{eqn:JRG}
\begin{split}
&R_{R/G}(x)\\
&=R_{SRH}(x)+R_{radiative}(x)+R_{Auger}(x)+ R_G(x).    
\end{split}
\end{equation}
%modeled via ABC equations~\cite{Piprek2010} involving three kinds of processes. 
The SRH recombination rate is given by Ref.~\cite{Piprek2010}  
\begin{equation} \label{eqn:A}
\begin{split}
&R_{SRH}(x) \\ &=\frac{N_n(x)N_p(x)-n_{intrinsic}^2}{(N_n(x)+n_{intrinsic})/A+(N_p(x)+n_{intrinsic})/A}
\end{split}
\end{equation}
with $A=2.6\times 10^6 s^{-1}$, the empirical parameter of the inverse recombination lifetime~\cite{Kim2007,Dmitriev1999,Olivier2017},  $n_{intrinsic}$ the intrinsic carrier density\cite{Schaffl2001} and $N_{n,p}$, the density of electrons (n) and holes (p).
We solve the radiative recombination rate by~\cite{Piprek2010}
\begin{equation} \label{eqn:B}
\begin{split}
& R_{radiative}(x) = B\cdot N_n(x)N_p(x) \\
\end{split}
\end{equation}
with the empirical parameter~\cite{Kim2007,Dmitriev1999,Olivier2017} $B=1.48\times10^{-11} cm^3 s^{-1}$. 
We determine the Auger recombination rate with~\cite{Piprek2010}
\begin{equation} \label{eqn:C}
\begin{split}
& R_{Auger}(x) = C\cdot \left( N_n(x)^2N_p(x)+N_p(x)^2N_n(x) \right) \\
\end{split}
\end{equation}
with the empirical parameter~\cite{Kim2007,Dmitriev1999,Olivier2017} $C=1.6\times10^{-30} cm^6 s^{-1}$. 
Note that we use the same ABC parameters for all NEGF and TCAD results literature. 
In most calculations, the generation current density $R_G(x)$ is set to 0. 
Only when explicitly mentioned that illumination is included, the generation current density $R_G$ in the well is determined by integrating photon numbers of the solar spectrum of energies larger than the bandgap of In$_{0.13}$Ga$_{0.87}$N. Outside the well, $R_G(x)$ is assumed to vanish
\begin{equation}
\label{Jq}
\begin{aligned}
&R_{G}(x)=
&\begin{cases}
R_G,   & x \in \mbox{InGaN} \\
0,     &\mbox{otherwise.} 
\end{cases}
%\Gamma_j(E) = 
\end{aligned}
\end{equation}

The performance of the two devices is solved in NEGF with the new B\"{u}ttiker probes.
The NEGF results are benchmarked against DD and quantum corrected DD model (SILVACO-ATLAS\cite{atlasweb}) results.

For all models, the spatially resolved recombination and generation rate ($R_{R/G}(x)$) is multiplied with the elementary charge and integrated along the total device to solve for the total recombination and recombination current density $J_{R/G}$. 
\begin{eqnarray}
\begin{aligned}
\label{Itotal}
&J_{R/G} = \int_0^L R_{R/G}(x) dx. \\
\end{aligned}
\end{eqnarray}
For each carrier type a current conservation law modified by the recombination and generation is fulfilled 
\begin{eqnarray}
\begin{aligned}
\label{kirch}
&J_{s,n} + J_{d,n} + J_{R/G} = 0   \\
&J_{s,p} + J_{d,p} - J_{R/G} = 0.   \\
%J_{total}=J_{thermionic}+J_{recombination}.
\end{aligned}
\end{eqnarray}
$J_{s/d,n}$ ($J_{s/d,p}$) defines the source/drain current density for electrons (holes). 
The total measurable current density at the source is given as
\begin{eqnarray}
\begin{aligned}
\label{Itotal}
&J_{total} = J_{s} =J_{s,n}+J_{s,p}.\\
\end{aligned}
\end{eqnarray}
An equivalent equation holds for the drain current density.
%Due to the built-in potential, $I_{thermionic}$ plays an important part at high $V_{sd}$ bias.
% $I_{recombination}$ dominates at low $V_{sd}$ bias due to the spatial overlapped of the electrons and holes. 
%end of move section

% \added{
% \begin{equation} \label{eqn:ABC}
% \begin{split}
% %&J_R(x) = J_{SRH}(x) + J_{radiative}(x) + J_{Auger}(x)\\
% & J_{SRH}(x) \\ &=\frac{N_n(x)N_p(x)-n_{intrinsic}^2}{(N_n(x)+n_{intrinsic})/A+(N_p(x)+n_{intrinsic})/A}  \\
% %&\tau_n = \tau_p = 1/A\\
% & J_{radiative}(x) = B\cdot N_n(x)N_p(x) \\
% & J_{Auger}(x) = C\cdot \left( N_n(x)^2N_p(x)+N_p(x)^2N_n(x) \right) \\
% & J_G=
% \end{split}
% \end{equation}
% }

%First, the implementation of B\"uttiker probe recombination and generation(BPRG) method  in NEMO5\cite{Steiger2011} is explained in details. Secondly, drift-diffusion(DD) based model from SILVACO-ATLAS is introduced. 
 
\subsection{NEGF with B\"{u}ttiker probes}

Electron and hole properties are solved within the NEGF method~\cite{Datta2000}. 
To limit the computational load, the devices are partitioned and Green's functions are solved recursively on the resulting slabs~\cite{Sadasivam2017}.
The retarded Green's function $G^R$ is solved by the Dyson equation\cite{Datta2000}
\begin{eqnarray}
\label{dyson_eqn}
G^{R} = [E-H-\Sigma_{S}^{R}-\Sigma_{D}^{R}-\Sigma_{BP}^{R}]^{-1}.
\end{eqnarray}
The lesser Green's function $G^<$ is given in the Keldysh equation
\begin{eqnarray}
\label{keldysh_eqn}
G^<=G^R (\Sigma^<_{S} + \Sigma^<_{D} +  \Sigma^<_{BP}   ) G^{R\dagger}.
\end{eqnarray}
All Green's functions and self-energies are matrices in the discretized positions space.
Their dependency on the in-plane momentum $k_{\parallel}$ and energy $E$ is omitted in Eqs.~(\ref{dyson_eqn}) and (\ref{keldysh_eqn}) for better readability. 
The source and drain contact self-energies are given by $\Sigma^{R,<}_{S}$ and $\Sigma^{R,<}_{D}$. 
Scattering of electrons and holes is included with B\"{u}ttiker probe self-energies ($\Sigma^{R,<}_{BP}$) ~\cite{Bttiker1986,Anantram2008,Sadasivam2017}.
Quantities such as the density of states, the particle density, the state occupancy and the current density can be deduced from the Green's functions as typical for the NEGF method~\cite{Datta2000,Kubis2011}.

%\subsection{Self energy}

%In Eq.(1) of Ref.~\cite{Vaitkus2017}, $\Sigma_{BP}$, the self-energy matrix, is implemented here as a diagonal matrix. 
The retarded B\"{u}ttiker probe combines all intra-band scattering processes, such as scattering on various phonons, impurities and electron-electron scattering into the empirical scattering parameter $\eta$~\cite{Geng2018}. 
To resemble the Urbach tail\cite{FranzUrbach1953,Sarangapani2018} in GaN, $\eta$ is exponentially decaying into the band gap.
For electrons in the conduction band, the retarded B\"{u}ttiker probe self-energy reads 
\begin{equation}
\label{self_energy_exp}
\begin{aligned}
&\Sigma_{BP,n/p}^{R}(x,x',k_{\parallel},E) = \delta(x-x') \\
\times  &\begin{cases}
    a \eta_n, & \text{if } E \geq E_c\left(x,k_{\parallel}\right)\\ 
    a \eta_n \cdot \exp \left(
    -\frac{E_c(x,k_{\parallel})-E}{\lambda}\right)
    , & \text{if } E_c\left(x,k_{\parallel}\right) > E \geq \frac{E_c+E_v}{2} \\
    a \eta_p \cdot \exp \left(
    \frac{E_v(x,k_{\parallel})-E}{\lambda}\right)
    , & \text{if }  \frac{E_c+E_v}{2} > E \geq E_v\left(x,k_{\parallel}\right) \\
    a \eta_p, & \text{if } E_v\left(x,k_{\parallel}\right)>E  .
\end{cases}
\end{aligned}
\end{equation}
In all NEGF calculations, the mesh size a is set to $0.259$~nm.

\begin{figure}[h]
 	\centering
 	\includegraphics[width=3.4in]{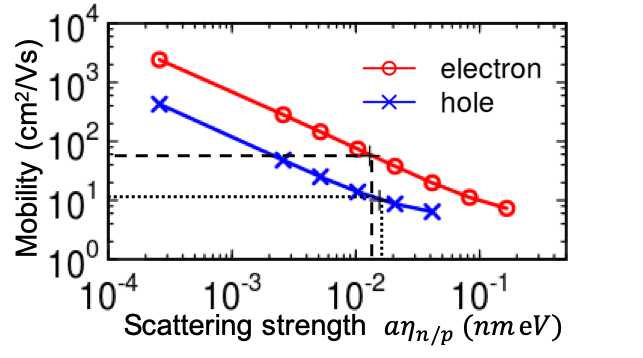}
 	\caption{NEGF predicted mobility of a homogeneous semiconductor with p and n doping density of $10^{20}/cm^3$ as a function of the product of $\eta_{n,p}$ of Eq.~(\ref{self_energy_exp}) with the mesh spacing $a$. } 
 	\label{eta_mobility} 
 	% folder:  \url{https://nanohub.org/groups/klimeck/svn/trunk/StudentData/kuangchungwang/sphinx_log/source/201711_buttikerprobe/4_flat_band_mobility}
 	% picture: eta_mobility.png 
 	% raw data: 
 	%1.eta_mobility_1e20input_refine_bandedge_uniform_fixedK_flat_contacteta.in
 	%2.eta_mobility_1e20input_refine_bandedge_uniform_fixedK_hole_flat_contacteta.in
 	% raw inputs:
 	%1. input_refine_bandedge_uniform_fixedK_hole_flat_contacteta.in 
 	%2. input_refine_bandedge_uniform_fixedK_flat_contacteta.in
 	%script to reproduce: 
 	%./submit_voltage.sh  sub    0 1e20 input_refine_bandedge_uniform_fixedK_flat_contacteta.in
 	%  ./submit_voltage.sh  sub     0 1e20 input_refine_bandedge_uniform_fixedK_hole_flat_contacteta.in
 	
 \end{figure}

The retarded B\"{u}ttiker probe for holes in the valence band has the same formula, but with valence band parameters.
Similar to Ref.~\onlinecite{Szabo2015}, the electron and hole mobility are deduced from the respective resistivity of n- or p-doped homogeneous material samples solved with NEGF and B\"{u}ttiker probes
\begin{eqnarray}
\label{resistivity_eqn}
\rho_{n,p} = \frac{dR^\Omega_{n,p}}{dL}  = \frac{1}{q(\tilde{\mu}_{n,p} N_{n,p})}.
%\label{mob_eqn}
%\rho }
\end{eqnarray}  % make R roman. 
Here, $R^\Omega$ refers to the resistance of GaN samples of length $L$, $R^\Omega(L)=V_{sd}/I(L)$.
The applied Fermi level difference of source and drain in the mobility calculation is set to $10$~meV.
$I(L)$ is the length dependent current density for electrons or holes and solved for $L=20$~nm and $L=25$~nm.
The empirical scattering parameter $\eta_{n}=0.05$~eV for electrons and $\eta_{p}=0.06$~eV holes are chosen such that the respective NEGF predicted mobility agrees with $\tilde{\mu}_e=56.88 cm^2/(Vs) $ and $\tilde{\mu}_h=10.0 cm^2/(Vs)$ (taken from Ref.~\onlinecite{Mnatsakanov}).
For completeness, Fig.\ref{eta_mobility} shows the GaN hole and electron mobility as a function of the respective $\eta$.
The band tail parameter $\lambda$ is chosen to be 55~meV for electrons according to Ref.~\onlinecite{Sarangapani2019}. 
The same value is assumed for holes. 

The "lesser than" B\"uttiker probe self-energy $\Sigma^{<}_{BP}$ is depending on the B\"uttiker probe Fermi-levels, $\mu_{n/p}(x)$ 
\begin{equation}
\label{Sig_in} 
\begin{aligned}
&\Sigma^<_{BP,n/p}(x,x',k_{\parallel},E) = \\
&\begin{cases}
-F\left( \mu_n(x),E \right) (\Sigma_R(x,x',k_{\parallel},E)-\Sigma_R^\dagger(x,x',k_{\parallel},E)), \\
 \text{          if }   E \geq \frac{E_c+E_v}{2} \\
-(1-F\left( \mu_p(x),E \right) (\Sigma_R(x,x',k_{\parallel},E)-\Sigma_R^\dagger(x,x',k_{\parallel},E)), \\
 \text{          if }    E < \frac{E_c+E_v}{2}.
\end{cases}
%\Gamma_j(E) = 
\end{aligned}
\end{equation}
Here, $F$ is the equilibrium Fermi distribution function.
In this work, $\mu_{n}(x)$ and $\mu_{p}(x)$ are solved iteratively to satisfy the overall current conservation
\begin{equation}
\begin{aligned}
\label{Current_conservation}
&R_{n}(x) =  R_{p}(x) =R_{R/G}(x)
%J_{SRH}(x)+J_{radiative}(x)+J_{Auger}(x)+ J_G(x). 
\end{aligned}
\end{equation}
$R_{n}$ and $R_{p}$ represent the net electron and hole current of the B\"uttiker probe at position $x$, respectively. 
In the state-of-the-art  B\"uttiker probe models, $\mu_{n}(x)$ and $\mu_{p}(x)$ are solved separately which ensures both, electron and hole current conservation individually.
Equation~(\ref{Current_conservation}) agrees with the common B\"uttiker probe model for the case of $R_{R/G}=0$.

The retarded source and drain self-energies $\Sigma^R_{S,D}$ are solved iteratively following Ref.~\cite{Sancho2000,Sancho1985}.
To guarantee smooth electron and hole transitions at the device/source and device/drain interfaces, the retarded B\"uttiker probe self-energy of Eq.~(\ref{self_energy_exp}) is included in the source and drain self-energy calculation~\cite{Kubis2009d,Miao2016}. 
The B\"uttiker probe and NEGF equations are iterated with the Poisson equation to achieve charge self-consistency. 
Piezoelectric and spontaneous polarizations are included as well following Ref.~\onlinecite{Geng2018}.  
%change bulk to contiunum 
\subsection{Drift diffusion models}
The purely semiclassical bulk drift-diffusion(DD) model~\cite{DeFalco2005} is applied on the homojunction pn diode with a mesh spacing of $0.1$~nm. 
In case of the pn diode including a quantum well, the DD model is augmented with quantum corrections for the bound states (DD+qwell) in the well region~\cite{atlasweb}. 
That heterojunction device is discretized with an adaptive real space mesh~\cite{atlasweb} with average mesh spacing of $0.16$~nm.
Depending on their energy and location, electrons and holes are separated into two groups, "bound states" and "bulk states".
For electronic energies below the barrier potentials ("bound states") at the GaN/InGaN interfaces, the Schr\"odinger equation is solved assuming the wave functions vanishing at the Schr\"odinger domain boundaries (i.e. Dirichlet boundary conditions). 
The solution domain of the Schr\"odinger equation is exceeding the InGaN quantum well by 10~nm in both directions to account for the wavefunction penetration into the barriers. 
Artificial bound states in GaN that would arise from the Dirichlet boundary conditions are avoided by limiting the lower bound of the GaN band edge to the barrier potential at the GaN/InGaN interfaces~\cite{atlasweb}.
For "bulk states", the DD equations are solved throughout the structure, but the InGaN band edge is shifted to the minimum of the barrier potentials at the unaltered GaN/InGaN interfaces~\cite{atlasweb,doi:10.1063/1.4808241}.
This ensures all "bulk states" do not face quantum well potential confinement.

Charge carriers of the two energy sets ("bound" and "bulk"), are coupled to each other by a capture-escape model~\cite{atlasweb}. 
Capture-escape rates for electrons and holes respectively are added to the DD continuity equations to allow transitions between "bound" and "bulk" particle groups~\cite{atlasweb}.
% \added{ which one should I add?
% \begin{equation} \label{eqn:capt}
% \begin{split}
% \frac { \partial \tilde { n } ^ { 2 D } } { \partial t } = \tilde { R } _ { c a p t , n } - \tilde { R } _ { r e c o m b } - \frac { 1 } { e } \nabla \tilde { J } _ { n } ^ { 2 D }
% \end{split}\\
% \frac { \partial n } { \partial t } = \frac { 1 } { q } \operatorname { div } \vec { J } _ { n } + G _ { n } - R _ { n }
% \end{equation}
% }
Recombination-generation mechanisms are included in the continuity equations as well (Eqs.~(\ref{eqn:A}-\ref{eqn:C})).
Detailed balance is ensured and all Fermi levels uniquely determined by using the same rates for carrier gain/loss in the "bound" and "bulk" groups and for generation and recombination, respectively. 
For charge self-consistency, the sum of the electron and hole "bound" and "bulk" density is iterated with the Poisson equation.

\section{Result}
Three application scenarios (a pn-diode, a pn-diode with a quantum well and an illuminated pn-diode with quantum well) are used to benchmark the B\"{u}ttiker probe model against the semiclassical and quantum corrected semiclassical models. 
The comparison shows the two methods agree very well in situations without pronounced quantum effects. 
The quantum corrections of the semiclassical model capture quantum effects but the results still deviate from those of a pure quantum mechanical treatment.

The pn diode is composed of 10~nm p type and 10~nm n type doped GaN with the doping concentration of $10^{20}/cm^3$ in each region.
Both regions are periodic in the transverse directions. 
The pn diode with the quantum well differs from the pn diode by a 3.0~nm thick intrinsic In$_{0.13}$Ga$_{0.87}$N quantum well layer. 
If not explicitly mentioned otherwise, the temperature is assumed to be $350$~K.
\subsection{GaN pn junction }
\begin{figure}[h]
 	\centering
 	 \includegraphics[width=3.4in]{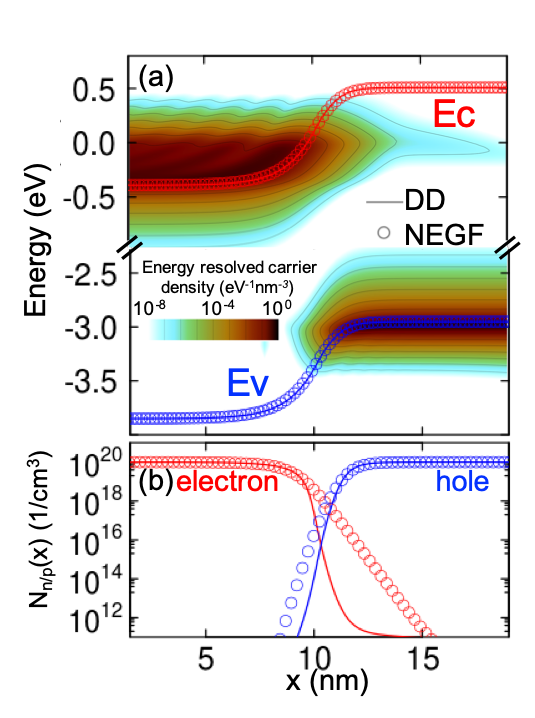}
 	 % 8 orders of 
 	\caption{The GaN pn diode described in the main text with a voltage of $V_{sd}=3.0V$ applied. (a) Conduction and valence band profiles solved in DD (lines) and NEGF with B\"{u}ttiker probes (symbols) along with contour plots of the energy resolved carrier densities at vanishing in-plane momentum. (b) Position resolved electron and hole  densities solved in DD (lines) and NEGF with B\"{u}ttiker probes (symbols).} 
 	\label{pn_charge} 
%   DD data:  	\url{ https://nanohub.org/groups/klimeck/svn/trunk/StudentData/kuangchungwang/sphinx_log/source/201711_buttikerprobe/silvaco_DD_comparison/pn_junction_updatedparam_update_mobi/well_Vd3.000000e+00_doping1e20pn1.in_jac_9_emesh_32_kmesh_20_kmax_0.5}
% NEGF data:   \url{https://nanohub.org/groups/klimeck/svn/trunk/StudentData/kuangchungwang/sphinx_log/source/201711_buttikerprobe/3_e_h_coupled_resonance_expdecay_semiclassical_GaN_EcEv_sym_exp_withpoisson_contacteta/Vd3.000000e+00_doping1e20input_refine_bandedge_uniform_fixedK_etaleads_urback.in_jac_9_emesh_128_kmesh_40_kmax_0.1_oneshot_}
% input deck :
% https://nanohub.org/groups/klimeck/svn/trunk/StudentData/kuangchungwang/sphinx_log/source/201711_buttikerprobe/3_e_h_coupled_resonance_expdecay_semiclassical_GaN_EcEv_sym_exp_withpoisson_contacteta/Vd3.000000e+00_doping1e20input_refine_bandedge_uniform_fixedK_etaleads_urback.in_jac_9_emesh_128_kmesh_40_kmax_0.1_oneshot_/temp_3.000000e+00_1e20input_refine_bandedge_uniform_fixedK_etaleads_urback.in_jac_9_emesh_128_kmesh_40_kmax_0.1.in
 \end{figure}

Figure~\ref{pn_charge}(a) shows the position resolved band edge of the GaN pn diode solved in DD and NEGF. 
Both results agree quantitatively.
The charge distributions of the two methods are depicted in Fig.~\ref{pn_charge}(b). 
The majority carrier in each region shows agreement between the two models while the minority charge from NEGF is around $10^3$ and $10^5$ times higher in the depletion region for electrons and hole, respectively. 
The minority carriers enter the oppositely doped area due to tunneling - as illustrated by the contour plot of the energy resolved carrier density of NEGF in Fig.~\ref{pn_charge}(a).
Note, the tunneling is longer ranged for conduction band states with their lighter effective mass than for the states in the valence band.
DD-based models do not capture this tunneling effect and the state-of-the-art quantum corrections do not apply to "bulk states".
The energy resolved density also illustrates Urbach tails with the additional density at energies below (above) the conduction (valence) band edge. 
It is worth to mention, decreasing $\lambda$ to $5$~meV in Eq.~(\ref{self_energy_exp}) reduced the Urbach tail and decreases the OFF-current by 10\% while the ON-current changes only marginally.

\begin{figure}[h]
 	\centering
 	\includegraphics[width=3.4in]{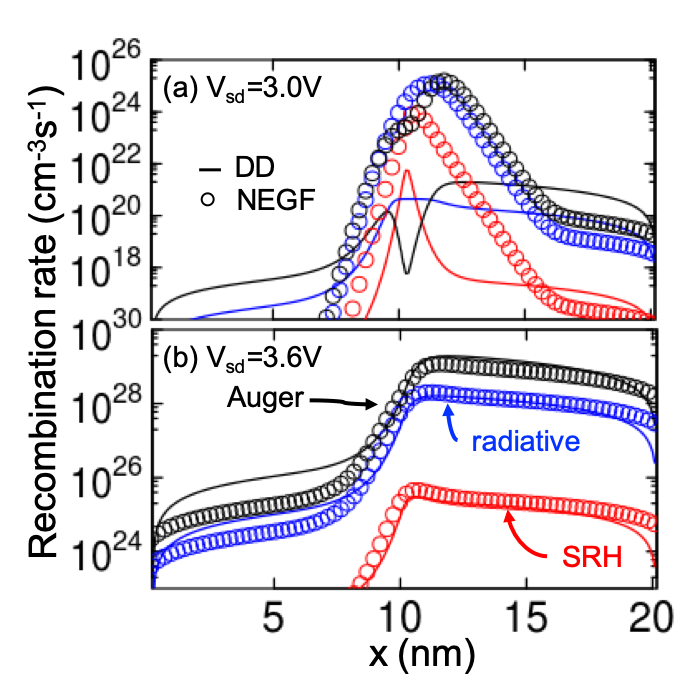}
 	\caption{ Spatially resolved recombination rates ($R_{SRH}(x)$, $R_{radiative}(x)$ and $ R_{Auger}(x)$) for the pn diode of Fig.~\ref{pn_charge} solved in NEGF with B\"{u}ttiker probes (symbols) and DD (lines) for applied voltages of V$_{sd}$=3.0~V (a) and 3.6~V (b). } 
 	%DD data for 3.0V:     \url{https://nanohub.org/groups/klimeck/svn/trunk/StudentData/kuangchungwang/sphinx_log/source/201711_buttikerprobe/silvaco_DD_comparison/pn_junction_updatedparam_update_mobi/well_Vd3.000000e+00_doping1e20pn1.in_jac_9_emesh_32_kmesh_20_kmax_0.5} 
 	%DD data for 3.6V \url{https://nanohub.org/groups/klimeck/svn/trunk/StudentData/kuangchungwang/sphinx_log/source/201711_buttikerprobe/silvaco_DD_comparison/pn_junction_updatedparam_update_mobi/well_Vd3.600000e+00_doping1e20pn1.in_jac_9_emesh_32_kmesh_20_kmax_0.5} 
 	%NEGF data for 3.0V:  \url{https://nanohub.org/groups/klimeck/svn/trunk/StudentData/kuangchungwang/sphinx_log/source/201711_buttikerprobe/3_e_h_coupled_resonance_expdecay_semiclassical_GaN_EcEv_sym_exp_withpoisson_contacteta/temp_3.000000e+00_1e20input_refine_bandedge_uniform_fixedK_etaleads_urback.in_jac_9_emesh_128_kmesh_40_kmax_0.1.in}
 	 %\url{https://nanohub.org/groups/klimeck/svn/trunk/StudentData/kuangchungwang/sphinx_log/source/201711_buttikerprobe/3_e_h_coupled_resonance_expdecay_semiclassical_GaN_EcEv_sym_exp_withpoisson_contacteta/Vd3.000000e+00_doping1e20input_refine_bandedge_uniform_fixedK_etaleads_urback.in_jac_9_emesh_128_kmesh_40_kmax_0.1trial_0}
 	%NEGF data for 3.6V: 
 	%\url{https://nanohub.org/groups/klimeck/svn/trunk/StudentData/kuangchungwang/sphinx_log/source/201711_buttikerprobe/3_e_h_coupled_resonance_expdecay_semiclassical_GaN_EcEv_sym_exp_withpoisson_contacteta/Vd3.600000e+00_doping1e20input_refine_bandedge_uniform_fixedK_etaleads_urback.in_jac_9_emesh_128_kmesh_40_kmax_0.1trial_0}
 	% \url{https://nanohub.org/groups/klimeck/svn/trunk/StudentData/kuangchungwang/sphinx_log/source/201711_buttikerprobe/3_e_h_coupled_resonance_expdecay_semiclassical_GaN_EcEv_sym_exp_withpoisson_contacteta/temp_3.600000e+00_1e20input_refine_bandedge_uniform_fixedK_etaleads_urback.in_jac_9_emesh_128_kmesh_40_kmax_0.1.in}
 	\label{pn_recomb} 
 \end{figure}

Differences in the density entail deviations of the recombination current density of the two models (see Eqs.~(\ref{eqn:A})-(\ref{eqn:C})).
This is illustrated in Fig.~\ref{pn_recomb} (a) which shows the various contributions to the position resolved recombination rate of DD and NEGF with B\"{u}ttiker probes for the situation in Fig.~\ref{pn_charge}(b).
For higher applied bias, the effective barrier between n- and p-doped region reduces.
Minority carrier tunneling becomes less relevant compared to the thermionic emission current \cite{Latreche2019}.
In consequence, the densities of DD and NEGF with B\"{u}ttiker probes at a higher voltages agree better and so do the recombination rate contributions in Fig.~\ref{pn_recomb}(b).

\begin{figure}[h]
 	\centering
 	\includegraphics[width=3.4in]{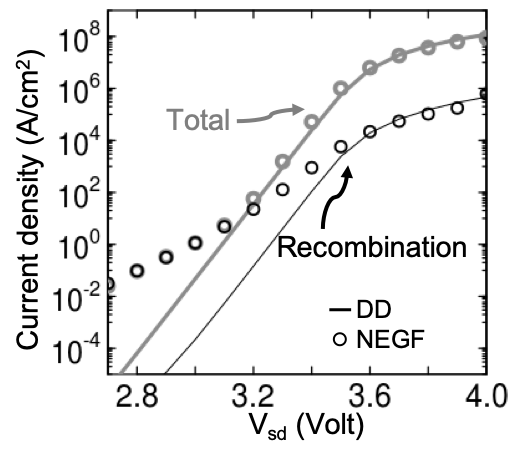}
 	\caption{Current-voltage characteristics of the pn diode of Fig.~\ref{pn_charge} predicted by NEGF with B\"{u}ttiker probes (symbols) and DD (lines). The total current density (gray) contains the contributions from the recombination current density (black).} 
     \label{pn_IV} 
 	% \url{https://nanohub.org/groups/klimeck/svn/trunk/StudentData/kuangchungwang/sphinx_log/source/201711_buttikerprobe/silvaco_DD_comparison/pn_junction_updatedparam_update_mobi/IVtotal_1e20pn1.in }
 	% NEGF data folder: \url{https://nanohub.org/groups/klimeck/svn/trunk/StudentData/kuangchungwang/sphinx_log/source/201711_buttikerprobe/3_e_h_coupled_resonance_expdecay_semiclassical_GaN_EcEv_sym_exp_withpoisson_contacteta}
 	% plot data file: 
 	 %\path{wellIVtotal_1e20input_refine_bandedge_uniform_fixedK_well_etaleads.in}
 	 %\path{wellIVrecomb_1e20input_refine_bandedge_uniform_fixedK_well_etaleads.in}
 	% raw data files: 
 	%current conservation loop: \path{Vd*_doping1e20input_refine_bandedge_uniform_fixedK_well_etaleads_urbach.in_jac_9_emesh_128_kmesh_40_kmax_0.1_oneshot_}
    %poisson calcuation: \path{Vd*_doping1e20input_refine_bandedge_uniform_fixedK_well_etaleads_urbach.in_jac_9_emesh_128_kmesh_40_kmax_0.1trial_0}
    % processing commands:   \path{overlord.sh}
 \end{figure}

The comparison of the total current density (Eq.~(\ref{Itotal})) predicted with DD and NEGF with B\"uttiker probes follows the same trend as can be seen in Fig.\ref{pn_IV}. 
Low voltages show pronounced deviations, whereas voltages above about $3.2$~V yield quantitive agreement in the total current density.
The recombination current density deviates until about $3.5$~V, but its relative contribution is insignificant for voltages above $3.2$~V.
Note that the slope of the IV-curve below $V_{sd}=$3.4V differs significantly between the two models due to the different treatment of tunneling.

 \subsection{InGaN quantum well embedded in GaN pn junction}
\begin{figure}[h]
 	\centering
 	\includegraphics[width=3.4in]{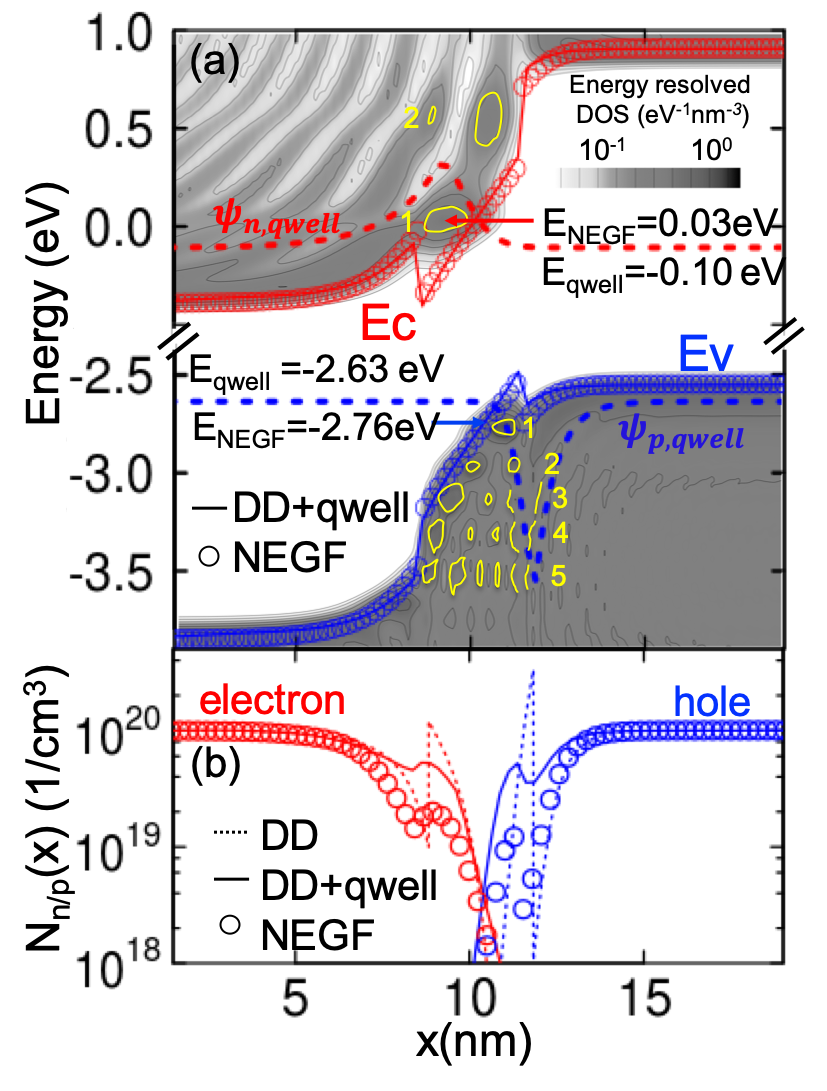}
 	\caption{The GaN pn diode with an InGaN quantum well described in the main text for an applied voltage of $V_{sd}=2.6V$. (a) Conduction and valence band profiles solved in DD+qwell (lines) and NEGF with B\"{u}ttiker probes (symbols) along with a contour plot of the energy resolved density of states at vanishing in-plane momentum. The dashed lines show the squared absolute value of the "bound" quantum well wave functions in the DD+qwell model. For reasons discussed in the main text, the ground state energies of DD+qwell and NEGF differ, and NEGF covers quasi-bound states in addition (indicated with yellow lines and labeled with numbers).
 	 (b) Position resolved electron and hole  densities solved in DD+qwell (lines) and NEGF with B\"{u}ttiker probes (symbols).
 	} 
 	%DD data Folder: 
 	%\url{https://nanohub.org/groups/klimeck/svn/trunk/StudentData/kuangchungwang/sphinx_log/source/201711_buttikerprobe/silvaco_DD_comparison/pn_junction_well_capture_escape_with_radiative_SRH_Auger_for_bound_states_with_pn_update_mobi/well_Vd2.600000e+00_doping1e20optimizeEkmesh_jac_9_emesh_32_kmesh_20_kmax_0.5}]
 	%Figure: \path{np_xnm_log2.png }   \path{Ec.png} \path{Ev.png}
 	%NEGF data Folder: 
 	%\url{https://nanohub.org/groups/klimeck/svn/trunk/StudentData/kuangchungwang/sphinx_log/source/201711_buttikerprobe/3_e_h_coupled_resonance_expdecay_semiclassical_GaN_EcEv_sym_exp_withpoisson_contacteta/Vd2.600000e+00_doping1e20input_refine_bandedge_uniform_fixedK_well_etaleads_urbach.in_jac_9_emesh_128_kmesh_40_kmax_0.1_oneshot_}
 	%input: \path{temp_2.600000e+00_1e20input_refine_bandedge_uniform_fixedK_well_etaleads_urbach.in_jac_9_emesh_128_kmesh_40_kmax_0.1.in}
 	%Contour raw data: 
 	%\path{bprgf_e_backward_solver_device_energy_resolved_k_integrated_dos.dat}
 	%\path{bprgf_h_backward_solver_device_energy_resolved_k_integrated_dos.dat}
 	%density data:  \path{debug_n.dat} \path{debug_p.dat}
 	\label{qw_band_charge} 
 	
 \end{figure}

When an In$_{0.13}$Ga$_{0.87}$N layer is added to the center of the GaN pn junction, a quantum well in the conduction and valence band forms (see Fig.~\ref{qw_band_charge}(a)).
Unaltered DD density results follow the band profiles\cite{DeFalco2005}.
Consequently, the unaltered DD calculations yield maxima in the carrier density close to the In$_{0.13}$Ga$_{0.87}$N/GaN interfaces (see Fig.~\ref{qw_band_charge}(b)).
In contrast, quantum corrected calculations cover quantum confinement effects with wavefunction maxima of "bound states" closer to the quantum well center (see dashed lines for conduction $\Psi_{n,qwell}$ and valence $\Psi_{p,qwell}$ band "bound states" in Fig.~\ref{qw_band_charge}(a)). 

Calculations of NEGF with B\"{u}ttiker probes do not distinguish between confined and continuum states but allow for smooth transitions between them.
This is illustrated with the contour plot of the energy and position resolved density of states (DOS) in Fig.~\ref{qw_band_charge}(a). %send tillmann the figure(Fig 5 )

Confined carriers in the quantum well extend into the continuum of states.
Continuum states are also modified by the interference effects at the quantum well boundaries. 
Higher order quantum well states with energies well beyond the barrier heights are still visible in the DOS continuum (yellow lines in Fig.~\ref{qw_band_charge}(a)).
These effects are missed in the quantum corrected DD model (DD+qwell) that separates "bound" and "continuum" spectra.
In detail, the conduction and valence band ground state energies predicted in NEGF are approximately $0.13$~eV higher than in DD+qwell. 
In consequence, the local density of the three models differ (see Fig.~\ref{qw_band_charge}(b)).

%With  B\"{u}ttiker probe model, the confined states's energy is extracted from the contour graph and shown to be different from the ones predicted from the Qwell due to the enforced Dirichlet B.C. vs. the entirely NEGF calculated result. The difference in the confined states energy and wavefulction results in different state filling and therefore explains the discrepancy in density in QW as in Fig. \ref{qw_band_charge}(b). 

\begin{figure}[h]
 	\centering
 	\includegraphics[width=3.4in]{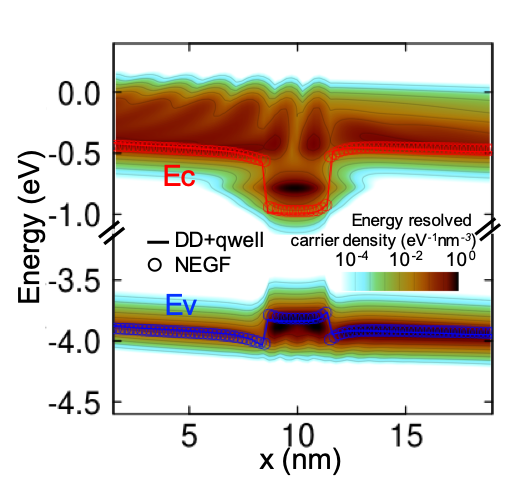}
 	\caption{The GaN pn diode with an InGaN quantum well of Fig.~\ref{qw_band_charge} with a voltage of $V_{sd} = 4.0V$ applied. Conduction and valence band profiles solved in DD+qwell (lines) and NEGF with B\"uttiker probes (symbols) are shown along with a contour plot of the energy resolved density of states at vanishing in-plane momentum solved in NEGF with B\"uttiker probes.
 	%(b) Carrier occupation from  B\"{u}ttiker probe and DD+qwell is compared at different locations in (a).
 	} %. smaller on the left, 
 	%two arrows , no shadow specify the dashed . 
 	% https://nanohub.org/groups/klimeck/svn/trunk/StudentData/kuangchungwang/sphinx_log/source/201711_buttikerprobe/3_e_h_coupled_resonance_expdecay_semiclassical_GaN_EcEv_sym_exp_withpoisson_contacteta/fermi.sh
 	%DD File : 
 	%\url{https://nanohub.org/groups/klimeck/svn/trunk/StudentData/kuangchungwang/sphinx_log/source/201711_buttikerprobe/silvaco_DD_comparison/pn_junction_well_capture_escape_with_radiative_SRH_Auger_for_bound_states_with_pn_update_mobi/well_Vd4.000000e+00_doping1e20optimizeEkmesh_jac_9_emesh_32_kmesh_20_kmax_0.5}
 	% \path{Ec.dat} \path{Ev.dat} 
 	%NEGF File:
 	%\url{https://nanohub.org/groups/klimeck/svn/trunk/StudentData/kuangchungwang/sphinx_log/source/201711_buttikerprobe/3_e_h_coupled_resonance_expdecay_semiclassical_GaN_EcEv_sym_exp_withpoisson_contacteta/Vd4.000000e+00_doping1e20input_refine_bandedge_uniform_fixedK_well_etaleads_urbach.in_jac_9_emesh_128_kmesh_40_kmax_0.1_oneshot_}
 	 %\path{bprgf_e_backward_solver_device_energy_resolved_k_integrated_density.dat}
 	 %\path{bprgf_h_backward_solver_device_energy_resolved_k_integrated_density.dat}
 	\label{qw_occu} 
 \end{figure}

\begin{figure}[h]
 	\centering
 	\includegraphics[width=3.4in]{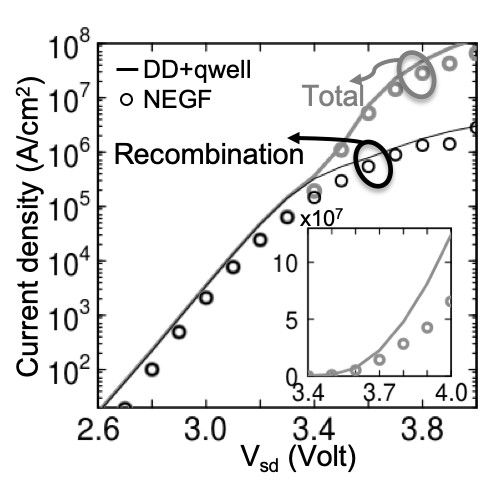}
 	\caption{Current-voltage characteristics of the GaN pn diode with a InGaN quantum well of Fig.~\ref{qw_band_charge} predicted by NEGF with B\"{u}ttiker probes (symbols) and DD+qwell (lines). The total current density (gray) contains the contributions from the recombination current density (black). The inset illustrates a difference of approximately 2x in the ON-current predicted with NEGF and DD.} 
 	\label{qw_IV} 
 	% figure file \url{https://nanohub.org/groups/klimeck/svn/trunk/StudentData/kuangchungwang/sphinx_log/source/201711_buttikerprobe/silvaco_DD_comparison/pn_junction_well_capture_escape_with_radiative_SRH_Auger_for_bound_states_with_pn_update_mobi/IVlog_eta0.005_comp.png}
 	% NEGF data folder: \url{https://nanohub.org/groups/klimeck/svn/trunk/StudentData/kuangchungwang/sphinx_log/source/201711_buttikerprobe/3_e_h_coupled_resonance_expdecay_semiclassical_GaN_EcEv_sym_exp_withpoisson_contacteta}
 	% plot data file: 
 	 %\path{wellIVtotal_1e20input_refine_bandedge_uniform_fixedK_well_etaleads.in}
 	 %\path{wellIVrecomb_1e20input_refine_bandedge_uniform_fixedK_well_etaleads.in}
 	% raw data files: 
 	%\path{Vd*_doping1e20input_refine_bandedge_uniform_fixedK_well_etaleads_urbach.in_jac_9_emesh_128_kmesh_40_kmax_0.1_oneshot_}
    %\path{Vd*_doping1e20input_refine_bandedge_uniform_fixedK_well_etaleads_urbach.in_jac_9_emesh_128_kmesh_40_kmax_0.1trial_0}
    % processing commands is in overlord.sh
 	% DD data:
 	%folder: \url{https://nanohub.org/groups/klimeck/svn/trunk/StudentData/kuangchungwang/sphinx_log/source/201711_buttikerprobe/silvaco_DD_comparison/pn_junction_well_capture_escape_with_radiative_SRH_Auger_for_bound_states_with_pn_update_mobi/}
 \end{figure}

% flip around the paragraph follow graph's order 
% story line: higher voltage shows clearly confined conduction band state
%still: signficant difference to DD+qwell, since continuum states modified in the presence of quantum well potential ("ribbles in EDEN").
%even bound state occupancy differs (b). 
For an applied voltage of V$_{sd}$=4.0V the quantum well ground states of conduction and valence bands are well confined (see Fig.~\ref{qw_occu}).
In this case, a distinction of "bound" and "continuum" states as done in DD+qwell is obvious.
%Most of the local occupancy function of NEGF~\cite{Kubis2011} agrees with local Fermi distributions (see Fig.~\ref{qw_occu}(b)). 
%Electrons that coherently propagate long distances cause a nonequilibrium distribution at the p-side of the device (at $x=15$~nm in Fig.~\ref{qw_occu}(b)).
%This effect is missed in DD models which yield equilibrium distribution functions throughout the device.
%in conjunction with the B\"{u}ttiker probe scattering yield differences in the electron and hole occupancy function~\cite{Kubis2011} as shown in Fig.\ref{qw_occu}(b). 
However, the energy resolved density in Fig.~\ref{qw_occu} for energies above the barrier potential (i.e. "continuum" states) still shows significant interference due to the potential change at the quantum well.
The interference pattern of the electron (hole) density are best visible in the n-doped (p-doped) region.
These quantum effects add resistance to the B\"{u}ttiker probe scattering.
Since the B\"{u}ttiker probe scattering strength was tuned to match the mobility assumed in the DD calculations, the total current density of the device in Fig.~\ref{qw_occu}  is predicted lower in NEGF than DD.
This is illustrated in Fig.~\ref{qw_IV} which shows the total ON current is about two times smaller in NEGF than DD.
Since the density of the confined state in the quantum well of NEGF calculations is smaller than in DD (see Fig.~\ref{qw_band_charge}b), the recombination current density of NEGF is smaller than in DD, too (see black lines and symbols Fig.~\ref{qw_IV}).

%It is worth to note the NEGF predicted occupancy contains nonequilibrium effects (best visible for electrons at $x=5.1$~nm), whereas carriers are forced to relax to a  local quasi Fermi level in all DD models. 

% arb units. 

\subsection{Illuminated InGaN quantum well embedded in GaN pn junction}
\begin{figure}[h]
 	\centering
 	\includegraphics[width=3.4in]{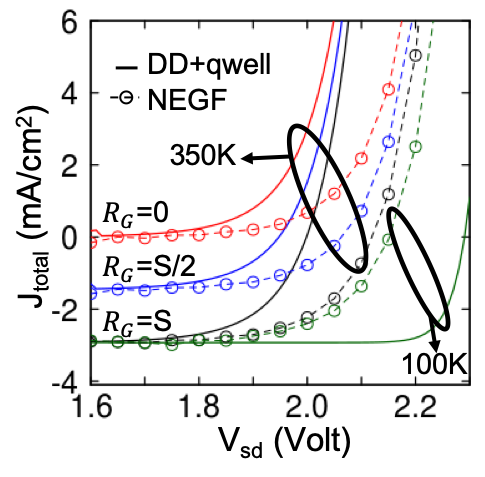}
 	\caption{Current-voltage characteristics of the GaN pn diode with a InGaN quantum well of Fig.~\ref{qw_band_charge} predicted by NEGF with B\"{u}ttiker probes (symbols) and DD+qwell (lines) with a finite electron-hole generation rate of S$=6.4\times 10^{22} cm^{-3}s^{-1}$ in the quantum well. The different quantum mechanical treatment of the bound and quasi-bound states of the two methods leads to  differences in the open-circuit voltage at 350~K and 100~K as detailed in the main text.} 
 	\label{qw_solar} 
 	%NEGF folder :
 	%\url{https://nanohub.org/groups/klimeck/svn/trunk/StudentData/kuangchungwang/sphinx_log/source/201711_buttikerprobe/3_e_h_coupled_resonance_expdecay_semiclassical_GaN_EcEv_sym_exp_withpoisson_solar_contacteta}
 	%input template: 
 	%\path{input_refine_bandedge_uniform_fixedK_well_contacteta_100k.in}
 	%\path{input_refine_bandedge_uniform_fixedK_well_contacteta.in}
 	%script: \path{overlord.sh} 
 	%figure:\path{IVnew350k_100knew.png}
 	%raw data: 
 	%\path{IV_350K IVsolar_half_350k IVsolar_350k IVsolar_100k well_new_IVtotal_1e20input_refine_bandedge_uniform_fixedK_well_contacteta.in_gen0 well_new_IVtotal_1e20input_refine_bandedge_uniform_fixedK_well_contacteta.in_gen1.325E-18 well_new_IVtotal_1e20input_refine_bandedge_uniform_fixedK_well_contacteta.in_gen2.65E-18 well_new_IVtotal_1e20input_refine_bandedge_uniform_fixedK_well_contacteta_100k.in_gen2.65E-18}
 	%DD folder :
 	%\url{https://nanohub.org/groups/klimeck/svn/trunk/StudentData/kuangchungwang/sphinx_log/source/201711_buttikerprobe/silvaco_DD_comparison/pn_junction_well_capture_escape_with_radiative_SRH_Auger_for_bound_states_with_pn_update_mobi_solarcell}
 	%
 \end{figure}

When various carrier generation rates due to solar light absorption are included in the calculations of the InGaN quantum well system of Figs.~(\ref{qw_band_charge})-(\ref{qw_IV}), the predicted current density is linearly shifted to negative values (see Fig.~\ref{qw_solar}) as common for solar cell operations~\cite{Tobnaghi2013}. 
The open circuit voltage for DD calculations at $350$~K temperature is smaller than in the NEGF case.
This is a result of the different quantum well densities and recombination rates discussed already in Figs.~\ref{qw_band_charge}(b) and \ref{qw_IV}.
With a lower temperature of $100$~K, the ground state energy of the quantum well turns out to be larger than the confinement potential and the DD model does not find any "bound states".
Then, the quantum well density of DD calculations is lower than in NEGF, and same holds for the recombination current density.
Therefore, at $100$~K, a higher open circuit voltage is observed in DD than in NEGF.

%\subimport*{paper_3_result/}{text.tex}

\section{Conclusion}
This work augmented the B\"{u}ttiker probe based scattering model in the nonequilibrium Green's function framework to efficiently model electron-hole recombination and generation processes.
Electrons and holes are modeled as separate particles that observe particle continuity equations with explicit particle creation and destruction rates.
The combined system of holes and electrons conserves the current.
These B\"{u}ttiker probes are applied on transport of electrons and holes in GaN pn-junctions with and without an embedded InGaN quantum well.
The results are benchmarked against Drift-Diffusion and quantum-corrected Drift-Diffusion calculations of Silvaco's TCAD tool Atlas~\cite{atlasweb}. 
The two methods agree quantitatively except for situations with pronounced carrier tunneling and interference. 
In particular, interference effects in the band continuum can create quasi-bound states that impact the optoelectronic device performance. 
In consequence and depending on the detailed device geometry, temperature and applied bias the open-circuit voltage of quantum well pn-junctions is underestimated or overestimated in state-of-art TCAD simulations.

\section*{Acknowledgment}
We acknowledge the Rosen Center for Advanced Computing at Purdue University for the use of their computing resources and technical support. 
The authors acknowledge the Texas Advanced Computing Center (TACC) at the University of Texas at Austin for providing high-performance computing resources. 
 \bibliographystyle{apsrev4-1k}
 %\bibliography{library.bib}
  \bibliography{references.bib}
  
% \onecolumngrid
% \section{ Reproduce the data :}
% \input{paper2_documentation.tex}

\end{document}